\documentclass{PoS}

\def\ie{{\it i.e.}}

\def\b{\beta}

\def\g{\gamma}
\def\h{\eta}

\def\j{\psi}

\def\p{\pi}                     
\def\t{\tau}

\def\z{\zeta}

\def\G{\Gamma}

\def\bj{\overline{\j}}
\def\bz{\overline{\z}}

\def\svev#1{\left\langle #1\right\rangle}       

\title{Running coupling from gluon exchange\\[1ex]
  in the Schr\"odinger functional}

\ShortTitle{Gluon-exchange coupling}

\author{\speaker{Yigal Shamir}\\
        Raymond and Beverly Sackler School of Physics and Astronomy\\
        Tel-Aviv University, Ramat~Aviv, 69978~Israel\\
        E-mail: \email{shamir@post.tau.ac.il}}

\abstract{
  I propose a new method to determine the running coupling in a
Schr\"odinger-functional setup.  The method utilizes the scattering amplitude
of massless fermions propagating between the time boundaries.
Preliminary tests show the statistical fluctuations of the new
observable to be about half those of the standard Schr\"odinger-functional
running coupling.}

\FullConference{ The XXIX International Symposium on Lattice Field Theory -
Lattice 2011\\
July 10-16, 2011\\
Squaw Valley, Lake Tahoe, California}

\begin{document}


\section{Introduction}
There is a growing body of non-QCD lattice work relevant to physics
beyond the Standard Model.  The vector-like theories under study
vary in their number of colors, their number of flavors,
and the group representation chosen for the fermions.

At present, the main interest in these models is as
candidate technicolor theories.  Phenomenological constraints \cite{TC}
require that a successful technicolor theory have a slowly running
coupling constant (``walking technicolor''),
as well as a large mass anomalous dimension.  A reliable determination of
these quantities requires nonperturbative lattice techniques.

The Schr\"odinger functional (SF) is a finite-volume setup especially suited
for this study. By suitably prescribing the spatial components
of the gauge field on the time boundaries, a classical, background
electric field is induced throughout the four-volume.
Measuring the response of the system to changes in the boundary values
then allows us to calculate the running coupling $g$ as a function of
the system's size $L$.

While providing a \textit{de-facto} standard method to measure
the running coupling, there are reasons to seek alternative
methods to determine the running coupling within the SF setup.
One reason is that the standard method relies on a very noisy observable,
while alternatives might be less so.
Also, an alternative method to determine the running coupling
will likely have different systematics, thereby providing
an important consistency check.  This is all the more important because
in theories with a slowly running coupling one is forced to use (relatively)
strong bare coupling in order to explore the vicinity of a (tentative)
infrared fixed point.

Another standard measurement in the SF setup is the extraction
of the mass anomalous dimension from the volume dependence
of $Z_P$, the renormalization constant of the pseudoscalar density.
This measurement turns out to be far less noisy than that
of the running coupling.

Motivated by this simple observation, I will explore the possibility
of extracting the running coupling from fermion correlation functions
that are not unrelated to those used in the determination of $Z_P$.
These correlation functions involve an ``ingoing'' and an ``outgoing''
fermion--antifermion pair.
The ingoing fermions have zero spatial momentum,
while the outgoing ones have nonzero momentum.
This kinematics enforces the exchange of a gluon between the two fermion lines,
and the result is that the correlation function is parametrically
of order $g^2$.

\section{The Schr\"odinger functional}

I will briefly describe the lattice implementation of the SF for two flavors
of Wilson fermions \cite{SF}.  It is assumed that the hopping parameter
has been tuned to its critical value by enforcing an axial Ward identity,
making the fermions massless.

The lattice is a four-dimensional mesh of $N^4$
sites, whose linear size is $L=Na$, where $a$ is the lattice spacing.
The spatial components $U_k=\exp(iaA_k)$
of the SU($N$) gauge field on the time boundaries
(at $t=0$ and $t=L$) are fixed to have commuting, spatially constant values.
Different values are chosen for the two boundary, with the effect
that $S_{cl}$, the minimum of the classical action,
describes a constant electric field throughout the bulk.  In lattice
perturbation theory, calculating the first quantum correction
to the effective action $\G$ gives
\begin{equation}
  \G = \left( \frac{1}{g_0^2}
           + \frac{b_1}{32\p^2} \log(L/a) \right) S_{cl}
  = \frac{1}{g^2(L)}\ S_{cl}
\end{equation}
where $g_0$ is the bare lattice coupling, and $b_1$ is the coefficient
of the one-loop beta function.  This result is true up to discretization
errors, \ie, up to corrections proportional to positive powers of the lattice
spacing $a$.  The boundary values of $A_k$ depend on the dimensionful parameter
$L$ only.  As a result,
$g(L)$ is the one-loop running coupling.  If we allow the boundary data
to depend in addition on a dimensionless parameter $\h$,
the $\h$-derivative of the effective action is an observable
that can be calculated using standard Monte-Carlo techniques.
The SF running coupling is thus defined nonperturbatively as
\begin{equation}
  \frac{K}{g_{SF}^2(L)} = \frac{\partial\G}{\partial\h}\bigg|_{\h=0} \;,
\qquad \qquad
  K = \frac{\partial S_{cl}}{\partial\h}\bigg|_{\h=0} \;.
\end{equation}
The minimal classical action $S_{cl}$, and hence $K$,
depend on $L/a$, but, as it turns out, the discretization
errors are of order $a^4$, making them practically negligible.

The pseudoscalar renormalization constant $Z_P=Z_P(L)$ is calculated in
the SF setup via \cite{ZP}\footnote{
  I omit normalization factors conventionally included in these definitions.
}
\begin{eqnarray}
  Z_P &=& f_P(L/2) / \sqrt{f_1} \,,
\label{ZP}\\
  f_P(t) &=& \sum_{\vec{x}} \svev{\bj(\vec{x},t)\g_5 \t_a \j(\vec{x},t) \;
                                \z \g_5 \t_a \z} \,,
\\
  f_1 &=& \svev{\bz' \g_5 \t_a \z' \; \bz \g_5 \t_a \z} \,.
\label{f1}
\end{eqnarray}
The Pauli matrices $\t_a$ act on the isospin index of the fermion field.
$\z$ and $\bz$ ($\z'$ and $\bz'$) are gauge invariant, zero-momentum
fermion and antifermion wall sources located near the $t=0$
($t=L$) boundaries \cite{ZP}.  The anticipated scaling behavior is
\begin{eqnarray}
   f_P(L/2) &\sim& L^3 Z_P(L) Z_\z^2(L) \,,
\label{scaleft}\\
   f_1 &\sim& L^6 Z_\z^4(L) \,,
\label{scalef1}
\end{eqnarray}
where $Z_\z$ is the wall-source renormalization factor.
The ratio~(\ref{ZP}) is independent of $Z_\z$,
thereby providing an acceptable prescription for $Z_P$.

\section{Running coupling from gluon exchange}

Let us turn our attention to the correlation function~(\ref{f1}) of the four
wall sources, used for normalization in the definition of $Z_P$.
We generalize this correlation function to
\begin{equation}
  f(\vec{n}) = \svev{\bz'(-\vec{n}) \g_5 \t_a \z'(\vec{n}) \;
                 \bz \g_5 \t_a \z} \,.
\label{mom}
\end{equation}
Here we have allowed the wall sources located near the $t=L$ boundary to have
nonzero momentum $\vec{p} = (2\p/L) \vec{n}$,
where the integer-valued vector $\vec{n}$ lives on the reciprocal lattice.
The wall sources near the $t=0$ boundary are kept at zero momentum.
Let us now consider the ratio
\begin{equation}
  R(\vec{n}) = \frac{f(\vec{n})}{f(\vec{0})} = \frac{f(\vec{n})}{f_1} \,,
\label{R}
\end{equation}
where we fix the components of $\vec{n}$ at small integers.
In the absence of any other infrared scale, the anticipated scaling is
\begin{equation}
   f(\vec{n}) \ \sim\ L^6 Z_\z^4(L) g^2(L) \,, \qquad  \vec{n} \ne \vec{0}\,.
\label{scalefp}
\end{equation}
In contrast with Eq.~(\ref{scalef1}),
momentum must now be transferred between the fermion lines.
This requires a gluon exchange, which costs a factor of $g^2$.
In the ratio~(\ref{R}), once again, the dependence on $Z_\z$ cancels out,
and the scaling is $R(\vec{n}) \sim g^2(L)$.  We may thus define
a \textit{gluon-exchange running coupling} via
\begin{equation}
  g_{GE}^2(\vec{n};L) = \frac{R(\vec{n};L)}{K(\vec{n})} \,,
\label{ge}
\end{equation}
where now $K(\vec{n})$ is the tree-level value of $R(\vec{n};L)$
in the limit $L/a\to\infty$.
Each momentum vector $\vec{n}$ amounts to a different scheme.
Standard arguments imply that $g_{SF}^2(L)$ as well as
$g_{GE}^2(\vec{n};L)$, for all $\vec{n}$, have the same two-loop
beta function.

\begin{figure}
\hspace{.17\textwidth}
\begin{picture}(60,100)(0,0)
\put(0,0){\includegraphics*[width=.6\textwidth]{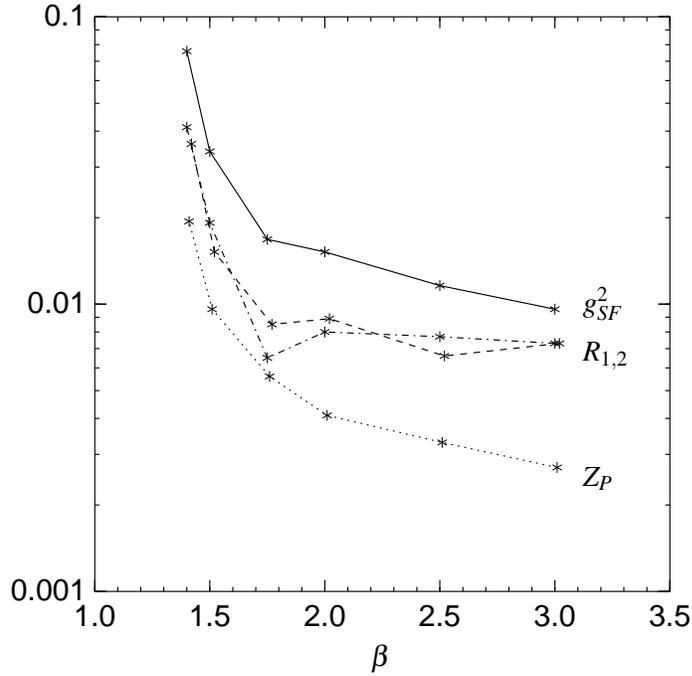}}
\put(58,-5){$\b$}
\put(92,24){$Z_P$}
\put(92,44){$R_{1,2}$}
\put(92,52){$g^2_{SF}$}
\end{picture}
\vspace*{3ex}
\caption{Relative error in several observables,
  plotted as a function of the bare coupling.
  Data points for $1/g^2_{SF}$, $R_1$, $R_2$, and $Z_P$,
  are connected by a solid, dashed-dot, dashed, and dotted line
  respectively. The $R_2$ points
  have been slightly shifted horizontally for better visibility.
  See text for further explanation. \label{rel-err}}
\end{figure}

A first look at the new observables is offered by Fig.~\ref{rel-err}.
This is an SU(2) gauge theory, with two Dirac fermions in the adjoint
representation \cite{it,BTY}.
An ensemble of 8000 configurations of volume $8^4$ was generated
for each of the six
bare-coupling values $\beta= 3.0$, 2.5, 2.0, 1.75, 1.5 and 1.4.
The lattice implementation is the same as in Ref.~\cite{BTY},
where also the values of $\kappa_c(\beta)$ may be found.
Fig.~\ref{rel-err} shows the relative error in several observables.
Two of them are the familiar $1/g_{SF}^2$ and $Z_P$.  The other two, $R_1$
and $R_2$, are linear combinations of $R(\vec{n})$ (Eq.~(\ref{R})),
for momentum vectors whose components are either $2\p/L$ or zero.
$R_1$ sums up $R(\vec{n})$
for the three possibilities where a single component of $\vec{n}$ is nonzero.
$R_2$ is similarly constructed, except that two momentum components
are nonzero.

As can be seen from the figure, the relative error in $Z_P$ is roughly
one fourth that of $1/g^2_{SF}$, while the relative errors in $R_1$
and $R_2$ are roughly in the middle.  This very preliminary result
suggests that the new observables indeed have smaller statistical fluctuations
than those of $1/g^2_{SF}$.

\section{Discussion}

In order to make use of the newly defined gluon-exchange coupling
it will be necessary to compute the relevant tree-level amplitudes,
$K(\vec{n})$, that enter as normalization constants
in the definition~(\ref{ge}).
Each $K(\vec{n})$ is the infinite-volume limit of the corresponding
finite-volume tree-level amplitudes.  A calculation of the tree-level
amplitudes at finite $L$ will also provide
important information about the discretization errors in the new observables.
Unlike in the very special case of $1/g^2_{SF}$, where, as mentioned above,
tree-level discretization errors go like $a^4$, for the gluon-exchange coupling
the discretization errors could be of order $a^2$ or even of order $a$.
It should be noted that, in the event that discretization errors
turn out to be relatively big for individual $R(\vec{n})$ amplitudes,
one may attempt to construct linear combinations of these amplitudes
that have smaller (tree level) discretization errors.  The question
will then be whether this can be done without degrading
the statistical quality of the observable.  Work on these questions
is underway.

\acknowledgments
I thank Tom DeGrand, Stefan Sint, Rainer Sommer, and Ben Svetitsky
for discussions and useful suggestions.  This research is supported by the
Israel Science Foundation under grant no. 423/09.

\end{document}